%% file: SP_paper.tex
\documentclass[11pt]{article}
\usepackage{amssymb,amsmath} 
\usepackage{graphicx}
\usepackage{epstopdf}
\usepackage[colorlinks=true, pdfstartview=FitV, linkcolor=blue, citecolor=blue, urlcolor=blue]{hyperref}

\input{GKmacros.tex}

\begin{document}

\title{Extended wave propagators as pulsed-beam communication channels\\\sv2
\small Plenary lecture at \sl Days on Diffraction \rm Conference, St. Petersburg, Russia\\
May 30-June 2, 2006. http://math.nw.ru/DD/}

\author{\bf Gerald Kaiser \rm\\
Physics Department, UT-Austin\\
and Center for Signals \& Waves\\
kaiser@wavelets.com\\  \hr{http://wavelets.com}
}

\maketitle

\begin{abstract}
Let $P(x_r-x_e)$ be the causal propagator for the wave equation, representing the signal received at the spacetime point $x_r$ due to an impulse emitted at the spacetime point $x_e$. Such processes are highly idealized since no signal can be emitted or received at a precise point in space and at a precise time. We propose a simple and compact model for \sl extended \rm emitters and receivers by continuing $P$ to an analytic function $\2P(z_r - z_e)$, where $z_e=x_e+iy_e$ represents a circular \sl pulsed-beam emitting antenna \rm centered at $x_e$ and radiating in the spatial direction of $y_e$ while $z_r=x_r-iy_r$ represents a circular \sl pulsed-beam receiving antenna \rm centered at $x_r$ and receiving \sl from \rm the spatial direction of $y_r$. The space components $\3y_e, \3y_r$ of $y_e, y_r$ give the spatial orientations and radii of the antennas, while their time components $s_e, s_r$ represent the time a signal takes to propagate along the antennas between the center and the boundary. The analytic propagator $\2P$ represents the transmission amplitude, forming a \sl communication channel. \rm Causality requires that the extension/orientation 4-vectors $y_e$ and $y_r$ belong to the future cone $V\6+$, so that $z_e$ and $z_r$ belong to the \sl future tube \rm and the \sl past tube \rm in complex spacetime, respectively. The imaginary ``retarded time'' $T_e=s_e-|\3y_e|/c$ represents the duration of the emitted pulse, and $T_r=s_r-|\3y_r|/c$ represents the integration time for the received pulse. The bandwidths of the antennas are $1/T_e$ and $1/T_r$, respectively.  The invariance of $\2P$ under imaginary spacetime translations ($z_e\to z_e+i\h,\ z_r\to z_r+i\h$) has nontrivial consequences.
\end{abstract}

\section{Introduction}

The purpose of this paper is to consider the following proposition:  \sl When extended analytically to complex spacetime in a sense to be made precise, retarded wave propagators describe the causal interaction between a source of finite spatial extent and a sink of finite spatial extent. The supports of the extended source and sink are give by the imaginary spacetime coordinates associated with the analytic continuation, up to an equivalence due to invariance under complex spacetime translations. \rm

We shall explain and prove this proposition in the case of the scalar wave equation. It also extends to Maxwell's equations \ci{K3,K4,K5} and possibly some other systems \ci{K90}.  Begin with the retarded wave propagator
\begin{align}\lab{prop}
P\0x=\frac{\d(t-r)}{4\p r}, \qq  x=x_r-x_e=(\3x,t)\in\rr4, \ r=|\3x|,
\end{align}
where we have set the constant propagation speed to $c=1$ by an appropriate choice of units. $P$ is a fundamental solution of the wave equation,
\begin{align*}
\Box P\0x\=(\pl_t^2-\D)P(\3x,t)=\d(\3x)\d\0t\=\d\0x.
\end{align*}
If we fix the \sl emission event \rm $x_e=(\3x_e, t_e)$ and vary the reception event $x_r=(\3x_r, t_r)$, then $P(x_r-x_e)$ may be regarded as a wave emanating from the emission point $\3x_e$ at time $t_e$, the wave front at any time $t_r$ being the sphere $|\3x_r-\3x_e|=t_r-t_e$ (with no wave when $t_r<t_e$). This gives the usual picture of an exploding wave propagating causally away from $\3x_e$. 

However, we can also fix the \sl reception event \rm $x_r$ and examine $P(x_r-x_e)$ as a function of the emission event $x_e$. Then the same causal propagator is seen to describe \sl imploding \rm wave fronts $|\3x_e-\3x_r|=-(t_e-t_r)$ at emission times $t_e<t_r$, absorbed at $\3x_r$ at $t_e=t_r$ with no waves at $t_e>t_r$. Although this situation is completely symmetrical to the one with fixed emitter, it causes a bit of discomfort because we never actually \sl see \rm waves imploding toward a point. But a little reflection shows that we do not ``see'' acoustic or electromagnetic waves exploding either (like waves on a pond). Rather, the behavior of $P(x_r-x_e)$ as a function of $x_r$ is measured by setting up a system of idealized ``test receivers'' (a spacetime version of point-receivers) at various $x_r$ and piecing together their outputs. When viewed like this, the imploding wave picture no longer seems in conflict with causality: fix a single point receiver at $x_r$ and set up a system of ``test emitters'' at various $x_e$. The resulting transmission is again $P(x_r-x_e)$, this time as a function of $x_e$. Note that applying of the wave operators $\Box_e, \Box_r$ in the emission and reception variables $x_e, x_r$, respectively, gives
\begin{align*}
\Box_eP(x_r-x_e)=\Box_rP(x_r-x_e)=\d\0x,\qq P(x_r-x_e)=0\ \hb{\rm for}\ t_e>t_r, 
\end{align*}
confirming that $P(x_r-x_e)$ is indeed an \sl advanced wave propagator \rm in $x_e$. 

The reason why fixing the receiver and varying the emitter seems unusual is that \sl the field concept itself is biased toward reception, as proved by its very terminology: the filed variable is the ''observation point,'' never the ``emission point.''  \rm The bias toward reception also manifests itself in the fact that while no one has a problem with emission from a \sl extended \rm source, the dual process of observing with an extended receiver is usually treated \sl indirectly \rm by resorting to abstract \sl reciprocity theorems \rm involving two anti-causal twists: first the emitter and receiver are exchanged, then a space-time reversal is performed (represented by a complex conjugation in the frequency domain). By contrast, the above version of reciprocity (focus on $x_e$ instead of $x_r$) is \sl direct \rm and never leaves the causal world. It automatically involves a space-time reversal since $x=x_r-x_e$.

We shall treat \sl extended \rm emitters and receivers in a symmetric manner by showing that an analytic extension of $P(x_r-x_e)$ to complex $x_e$ and $x_r$ represents emission \sl and \rm reception by extended objects that, in fact, resemble \sl antenna dishes \rm whose sizes and orientations are given by the values of the imaginary spacetime variables attached to $x_e$ and $x_r$, respectively.

\section{The extended wave propagator}

There is, of course, no way to analytically continue the numerator $\d(t-r)$ of 
\eq{prop}. But it follows from the Sochocki-Plemelj formula that
\begin{align}\lab{bdy}
\d\0t=\frac1{2i\p}\lim_{s\to 0\9+}\lp\frac1{t-is}-\frac1{t+is}\rp,
\end{align}
hence $P(\3x, t)$ is the jump in boundary values of the \sl Cauchy kernel \rm
\begin{align}\lab{Cauchy}
\2\d(t-is)\=\frac1{2\p i(t-is)}
\end{align}
going from the lower complex time half-plane to the upper half-plane. Thus we begin by extending $P\0x$ to complex time, leaving space real for now:
\begin{align}\lab{prop2}
\2P(\3x,\t)\=\frac{\2\d(\t-r)}{4\p r}=\frac1{8i\p^2 r(\t-r)},\qq \t=t-is.
\end{align}
Next, we complexify space by replacing the Euclidean distance $r=|\3x|$ by the \sl complex distance \rm \ci{K0}
\begin{align}
\2r(\3z)&=\sr{\3z^2}\=\sr{\3z\cdot\3z},\qqq \3z=\3x-i\3y\in\cc3\nt\\
&=\sr{r^2-a^2-2i ax_3},\qq a\=|\3y|, \qq x_3\=\3x\cdot\bh y,\lab{2r}
\end{align}
where $\bh y=\3y/a$ and it is assumed that $a>0$. $\2r$ is an analytic but double-valued function in $\cc3$ which will be restricted to a double-valued function in $\rr3$ by fixing $\3y\ne\30$. Substituting $r\to\2r$ in \eq{prop2}, we obtain the full extension of $P$ to complex spacetime:
\begin{align}\lab{prop3}
\2P\0z\=\2P(\3z,\t)\=\frac{\2\d(\t-\2r)}{4\p \2r}=\frac1{8i\p^2 \2r(\t-\2r)},\qq z=(\3z,\t)\in\cc4.
\end{align}
This will be the main object of our attention. We will show that while the \sl boundary values \rm of $\2P$ in the sense of \eq{bdy} describe idealized emission and reception \sl events, \rm the \sl interior values \rm of $\2P(z_r-z_e)$ describe more realistic processes representing emission from and reception in \sl extended \rm spacetime regions surrounding the events $x_e=\re z_e$ and $x_r=\re z_r$. The imaginary spacetime variables $y_e\=\im z_e$ and $y_r\=-\im z_r$ will be seen to determine the \sl extent \rm of the spacetime regions where the emission and reception occur. They effectively give the \sl scales \rm (durations and source dimensions) in space and time for the emission and reception processes, subject to equivalence under complex spacetime translations $z_e'=z_e+\z,\ z_r'=z_r+\z$.

\section{Interpretation of $\2P$ as a pulsed beam}

To interpret the extended propagator $\2P$, we begin by studying the complex distance $\2r$. Writing
\begin{align}\lab{pq0}
\2r=p(\3x)-iq(\3x),
\end{align}
where the dependence on the fixed imaginary variables $\3y=-\im\3z$ is suppressed, so \eq{2r} gives
\begin{align}\lab{pq}
p^2-q^2=r^2-a^2\ \hb{\rm and}\   pq=a x_3.
\end{align}
Hence the cylindrical coordinate $\r$ orthogonal to $\3y$ is given by
\begin{align}\lab{rho}
\r^2=r^2-x_3^2=a^2+p^2-q^2-p^2q^2/a^2=\frac{(a^2+p^2)(a^2-q^2)}{a^2}.
\end{align}
This shows that $q^2\le a^2$, with equality only on the $x_3$ axis, and 
\begin{align*}
p^2=r^2-(a^2-q^2)\le r^2,
\end{align*}
with equality again only on the $x_3$ axis. Thus $\2r$ satisfies the bounds
\begin{align*}
|p|=|\re\2r(\3z)|\le |\re\3z|=r\ \hb{\rm and}\ |q|= |\im\2r(\3z)|\le |\im\3z|=a,
\end{align*}
with equality if and only if $\3x$ is parallel to $\3y$. Furthermore, \eq{rho} and \eq{pq} give
\begin{align}\lab{os}
\frac{\r^2}{a^2+p^2}+\frac{x_3^2}{p^2}=1,\qqq 
\frac{\r^2}{a^2-q^2}-\frac{x_3^2}{q^2}=1,
\end{align}
showing that the level surfaces of $p^2$ are \sl oblate spheroids \rm  $\5E_p$, and those $q^2$ are \sl one-sheeted hyperboloids \rm $\5H_q$. These are easily proved to be orthogonal families sharing the circle
\begin{align}\lab{C}
\5C=\{\3x\in\rr3|\ r=a,\ x_3=0\}
\end{align}
as a common \sl focal set. \rm Since $\5C$ is precisely the set in $\rr3$ where $\2r=0$, it is also the set of \sl branch points \rm of $\2r$ in $\rr3$ (for given $\3y\ne\30$). When $\2r$ is continued analytically within $\rr3$ from any point $\3x\notin\5C$ around a simple closed loop threading $\5C$, it changes sign due to the double-valuedness of the complex square root. To make $\2r$ a single-valued function in $\rr3$, we choose a \sl branch cut \rm $\5B$ consisting of a \sl membrane \rm spanning $\5C$,
\begin{align*}
\pl\5B=\5C,
\end{align*}
and declare that $\2r$ changes sign upon crossing $\5B$, so that continuation along a closed loop now returns $\2r$ to its original value. The simplest choice of branch cut is the disk
\begin{align*}
\5E_0=\{\3x\in\rr3|\ p=0\}=\{\3x\in\rr3|\ r\le a,\  x_3=0\},
\end{align*}
which is the degenerate limit of $\5E_p$ as $p\to 0$. We choose the branch of $\2r$ with $p\ge 0$, so that in the far zone \eq{2r} gives
\begin{align*}
r\gg a\imp \2r\app r-ia\cos\q,\qqq \cos\q\=\bh x\cdot\bh y.
\end{align*}
Hence \eq{prop3} gives the far-zone expression
\begin{align}\lab{far}
r\gg a\imp \2P\app \frac1{8i\p^2r\{(t-r)-i(s-a\cos\q)\}}
\end{align}
which is seen to be a \sl pulsed beam \rm of $\q$-dependent duration
\begin{align}\lab{T}
T\0\q=s-a\cos\q
\end{align}
propagating in the direction of $\3y$ ($\q=0$) whose peak value occurs at $t=r$:
\begin{align}\lab{peak}
\2P_{\rm peak}\app \frac1{8\p^2r(s-a\cos\q)}=\frac1{8\p^2 r T\0\q}.
\end{align}
To avoid singularities away from the source, we must require that the imaginary spacetime coordinates  satisfy the constraint
\begin{align}\lab{sa}
s>a\=|\3y|, \ \hb{\rm \ie}\  y=(\3y, s)\in V\6+\=\{(\3x,t)\in\rr4|\ |\3x|<t\}.
\end{align}
That is, $y$ must belong to the \sl future cone \rm $V\6+$ of spacetime, or $z=x-iy$ must belong to the \sl past tube, \rm also known as the \sl forward tube \rm in quantum field theory \ci{SW64}. We postulate that \sl $s$ is the time required for a signal to propagate along the antenna between the center and the boundary, \rm so that the condition $s>a\=a/c$ is simply a statement of \sl causality. \rm 
From \eq{peak} we find that the \sl far-field radiation pattern \rm is
\begin{align*}
\5F\0\q=\frac1{8\p^2(s-a\cos\q)},
\end{align*}
which describes an \sl ellipse \rm of eccentricity $a/s$ with one focal point at the origin. Thus $\5F$ peaks sharply around $\q=0$ as $s\to a\9+$.  Note that $\2P$ has \sl no sidelobes. \rm

\section{General pulsed-beam wavelets and their sources}

The pulsed beam $\2P$ decays slowly in time and angle ($\5O(t\inv)$ and $\5O((s-a\cos\q)\inv)$, respectively) because these decays come from the Cauchy kernel \eq{Cauchy}. Faster decays in both time and angle are obtained by convolving $\2P$ with distributional signals. Choosing a \sl driving signal \rm $g_0\0t$, we define general \sl pulsed-beam wavelet \rm by
\begin{align}\lab{wav}
W(\3z, \t)=\ir dt'\ g_0(t')\2P(\3z, \t-t')=\frac{g(\t-\2r)}{4\p\2r}
\end{align}
where $g$ is an \sl analytic signal \rm of $g_0$, defined by
\begin{align}\lab{analsig}
g\0\t=\frac1{2i\p}\ir\frac{g_0(t')\, dt'}{\t-t'},\qq \t=t-is.
\end{align}
$W$ is interpreted as the signal \sl  broadcast \rm by the extended source using $g_0$ as input.  $g_0\0t$ should have reasonable decay so that the integral converges, but it need not be analytic. Then $g\0\t$ is analytic  everywhere except for the support of $g_0$:
\begin{align*}
\7\pl_\t g\0\t=0\qq\forall \t\notin\supp g_0\subset\4R.
\end{align*}
In particular, $g$ is analytic in the upper and lower half-planes, given there by the positive and negative frequency parts of $g_0$, respectively:
\begin{align}\lab{gg}
g(t-is)=-\frac{\sgn s}{2\p}\ir d\o\,\Q(\o s)e^{-i\o(t-is)}\1g_0\0\o,
\end{align}
where $\1g_0$ is the Fourier transform of $g_0$ and $\Q$ is the Heaviside function. (This is a special case of the multidimensional \sl analytic-signal transform \rm  
\ci{K90, K94,K3}.) If $g_0\0t$ vanishes on any open set, the analytic functions $g(\t)$ in the lower and upper half-planes are analytic continuations of one another. It is possible to achieve excellent localization in both time and angle, still without sidelobes, by choosing an appropriate \sl distribution \rm for $g_0$. For example,
\begin{align*}
g_0\0t=\pl_t^n\d\0t \imp g\0\t=\pl_t^n\2\d\0\t=\frac{(-1)^n n!}{2i\p\t^{n+1}}.
\end{align*}

The pulsed-beam wavelet \eq{wav} reveals the mathematical nature of the extension $P\to\2P$. Fixing any vector $y\in V\6+$, the jump discontinuity of $W(x-iy)$ across real spacetime $\rr4$ is
\begin{align*}
\lim_{\e\to {0\9+}}\LB W(x+i\e y)-W(x-i\e y)\RB=\frac{g_0(t-r)}{4\p r}\,.
\end{align*} 
The right side is a \sl hyperfunction, \rm a distribution in $\rr4$ representing the jump in boundary values of a holomorphic function in a complex spacetime domain \ci{I92, K88}, in this case the jump from the past tube to the future tube. Physcially, it represents the 
wave caused by broadcasting the original driving signal from an \sl idealized \rm point source:
\begin{align*}
S_0(\3x,t)\=\Box \lp\frac{g_0(t-r)}{4\p r}\rp=g_0\0t\d(\3x).
\end{align*}
Thus we define the \sl extended \rm source of $W\0z$ by applying the \sl real \rm wave operator $\Box_x$ to $W(x-iy)$:
\begin{align}\lab{Sz}
S(x-iy)\=\Box_x W(x-iy),
\end{align}
where the subscript is a reminder that we are differentiating only with respect to $x$ since the imaginary source coordinates are constant. (They characterize the source's radius $a$, orientation $\bh y$, and duration $T\00=s-a$ along the beam axis.) Note that $W(x-iy)$ is singular along the branch circle $\5C$ (where $\2r=0$) and discontinuous across the branch cut $\5B$ (where $\2r$ changes sign). Hence the wave operator $\Box_x$ must be applied in a \sl distributional \rm sense \ci{K0, K3}. A \sl formal \rm differentiation of $W$ gives $S=0$, which shows that $S$ is actually a distribution supported on the singularities of $W$, which form the \sl world-tube \rm in spacetime swept out by the branch cut $\5B\subset\rr3$. The behavior in time is analytic because $\t-\2r$ is bounded away from the real axis as long as $y\in V\6+$. So the important fact for us is that \sl at any time $t$, the spatial support of the source $S$ is the branch cut $\5B$. \rm  The distributional source $S$ and its regularized versions (equivalent Huygens sources supported on surfaces surrounding $\5B$) have been computed in \ci{K3}, both in the spacetime and the Fourier (wavenumber-frequency) domains. 

The physical significance of the analytic signal $g(\t-\2r)$ is that an extended source cannot radiate an arbitrary signal with infinite accuracy because waves from different parts of the source interfere. The factor $e^{-\o s}$ in \eq{gg} shows that $g(t-is)$ can faithfully reproduce time variations only up to the \sl time scale \rm $s$, and similarly $g(\t-\2r)$ can reproduce time variations only up to $\5O(s-q)$. This gives a physical interpretation of $W$ and shows that the imaginary spacetime vector $y=(\3y, s)$ is indeed a multidimensional version of the \sl scale parameter \rm in ordinary wavelet analysis \ci{K94}.

\section{Pulsed-beam wavelet communication channels}

Our main goal here is to show that the pulsed-beam wavelets $W(x-iy)$ form natural \sl communication channels. \rm For this purpose, write the complex spacetime variables as
\begin{align}\lab{zz}
z\=x-iy=z_r-z_e\,,\ \hb{\rm where}\ z_e=x_e+iy_e\ \hb{\rm and}\ z_r=x_r-iy_r\,,  
\end{align}
with
\begin{align}\lab{yy}
y_e=(\3y_e, s_e)\in V\6+\ \hb{\rm and }\  y_r=(\3y_r, s_r)\in V\6+\,,
\end{align}
that is,
\begin{align}\lab{sese}
s_e>a_e\=|\3y_e|\ \hb{\rm and}\  s_r>a_r\=|\3y_r|.
\end{align}
Again, these are merely statements of \sl causality \rm if we postulate that $s_e$ is the time it takes a signal to propagate along the emitting antenna from the center to the boundary and $s_r$ is the time it takes a signal to propagate along the receiving antenna from the boundary to the center. The  opposite signs in $z_e$ and $z_r$ are \sl mathematically \rm necessary to ensure that \begin{align*}
z=(x_r-iy_r)-(x_e+iy_e)=(x_r-x_e)-i(y_r+y_e)
\end{align*}
belongs to the past tube (as it must) because $V\6+$ is convex. The \sl physical \rm interpretation of the sign change is that while extended \sl emission \rm sources are parameterized by the \sl future tube \rm $x_e+iy_e$, extended \sl reception \rm sources are parameterized by the \sl past tube \rm $x_r-iy_r$. (Roughly, emission is to the future while reception is from the past!) Though we shall not justify the definitions \eq{zz} further here, we now explore their consequences and show that they make complete physical sense. We shall interpret $W(z_r-z_e)$ as a \sl transmission amplitude \rm for a pulsed-beam wavelet emitted by an extended source centered at $x_e$ and characterized by $y_e$, then received by an extended source centered at $x_r$ and characterized by $y_r$. A given pair of source parameters $z_e$ and $z_r$ thus represent the communication channel depicted in Figure 1.
\begin{figure}[htbp]
\begin{center}
\includegraphics[width=3.8 in]{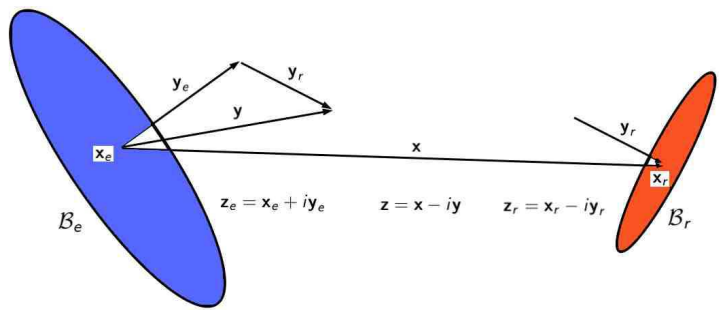}
\caption{$W(z_r-z_e)$ as a communication channel.}
\label{fig1}
\end{center}
\end{figure}
\noindent
We have sketched the emitting branch cut $\5B_e$, centered at $\3x_e$ with radius $a_e$ and orientation $\bh y_e$, and the receiving branch cut $\5B_r$, centered at $\3x_r$ with radius $a_r$ and orientation $\bh y_r$. Note that the vector $\3y_e$ points \sl out of \rm $\5B_e$ while $\3y_r$ points \sl into \rm $\5B_r$. This is a \sl graphical \rm explanation of the sign difference in \eq{zz}. To show that Figure 1 makes physical sense, note from \eq{gg} that $g$ decays with increasing distance from the real axis. Hence we need to minimize $|\!\im(\t-\2r)|=s-q$. This occurs when $q=a$, which implies $\3x$ is parallel to $\3y$. But
\begin{align}\lab{TT}
s-a=s_r+s_e-|\3y_r+\3y_e|\ge (s_r-|\3y_r|)+(s_e-|\3y_e|)
\end{align}
by the triangle inequality, and both terms on the right are positive by \eq{sese}. Therefore, $s-q$ is maximized, for given $a_e$ and $a_r$, when $\3x, \3y_e$ and $\3y_r$ are all \sl parallel. \rm That is, \sl  the gain is maximized when the two antennas are in a line-of-sight configuration. \rm Thus our \sl Ansatz \rm \eq{zz} about the dual complex structures $z_e=x_e+iy_e$ and $z_r=x_r-iy_r$ of the emitting and receiving sources leads to an obviously valid conclusion. (It would lead to the opposite conclusion had we set $z_r=x_r+iy_r$, which would also introduce unwanted singularities into the wavelet $W(z_r-z_e)$ since $y_r-y_e\notin V\6+$.)

Note that $W$ is invariant under \sl complex spacetime translations: \rm
\begin{align*}
z_e'=z_e+\z,\ z_r'=z_r+\z\imp z'=z_r'-z_e'=z,\qq \z=\x+i\h\in\cc4,
\end{align*}
which gives the new source vectors
\begin{align}
x_e'&=x_e+\x,\qq x_r'=x_r+\x,\qq x'=x_r'-x_e'=x\nt\\
y_e'&=y_e+\h,\qq y_r'=y_r-\h,\qq y'=y_r'+y_e'=y.\lab{invar}
\end{align}
(The imaginary translation vector  $\h$ must be sufficiently small that $y_e'$ and $y_r'$ remain in the future cone.) Invariance under \sl real \rm spacetime translation is nothing new, but its extension to \sl imaginary \rm translations  is new and interesting. Equation \eq{invar} states that the emitter-receiver pairs $(z_e, z_r)$ and $(z_e', z_r')$ form \sl equivalent channels. \rm This is far from obvious, though it makes intuitive sense. If we do \sl not \rm use a line-of-sight configuration for the emitter and receiver, we must increase the aperture sizes $a_e=|\3y_e|$ and  $a_r=|\3y_r|$ to obtain the same transmission quality. In particular, consider the following two ``extreme'' channels related to the channel depicted in Figure 1, which we will call channel A:

\begin{enumerate}

\item Channel $B$, equivalent to $A$ but with an idealized event receiver:
\begin{align*}
\z=iy_r\imp y_e'=y_e+y_r,\ \ y_r'=0.
\end{align*}
\item Channel $C$, equivalent to $A$ and $B$ but with an idealized event emitter:
\begin{align*}
\z=-iy_e\imp y_e'=0,\ \ y_r'=y_e+y_r.
\end{align*}
\end{enumerate}
\begin{figure}[htbp]
\begin{center}
\includegraphics[width=3.8 in]{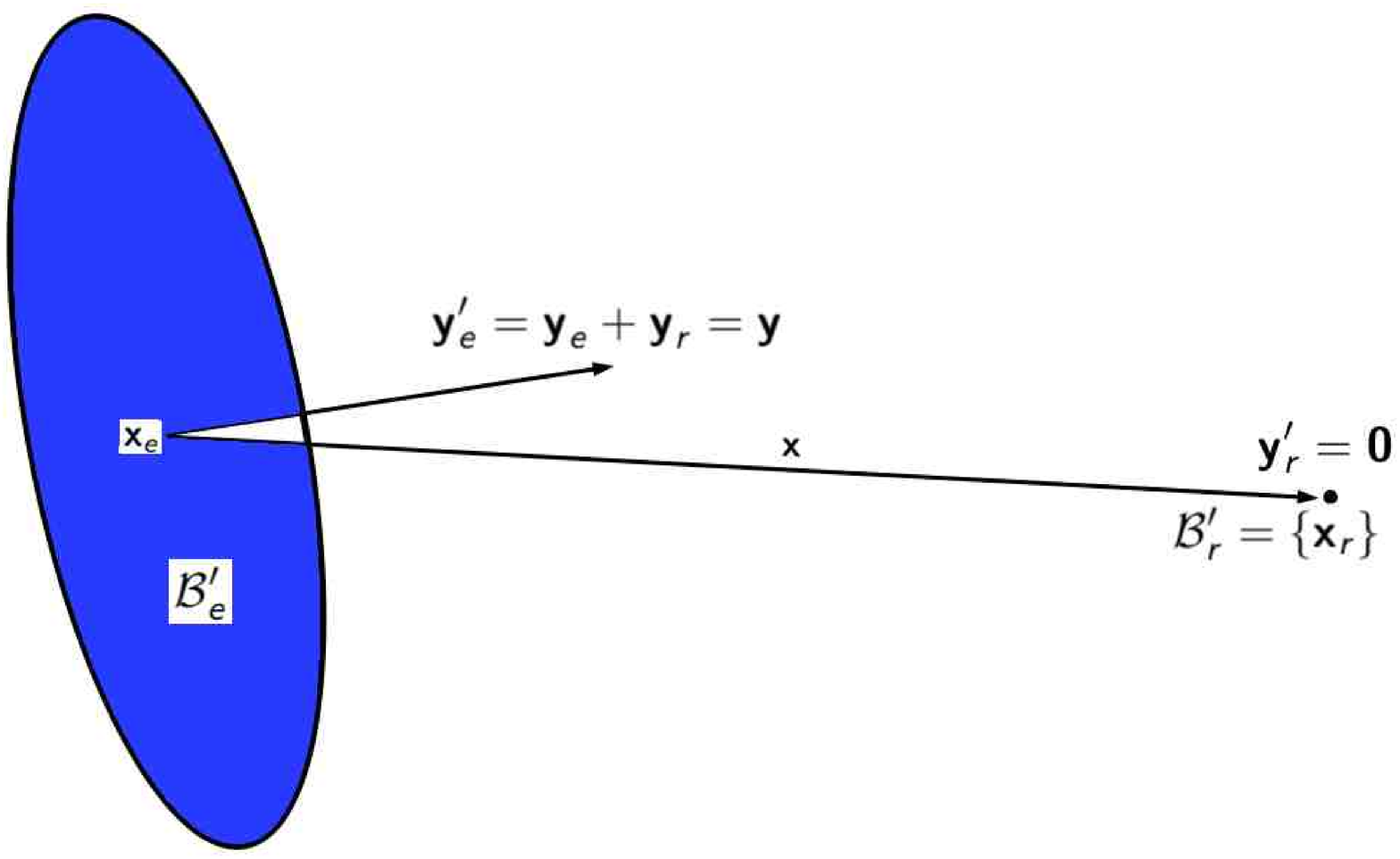}\\
\includegraphics[width=3.8 in]{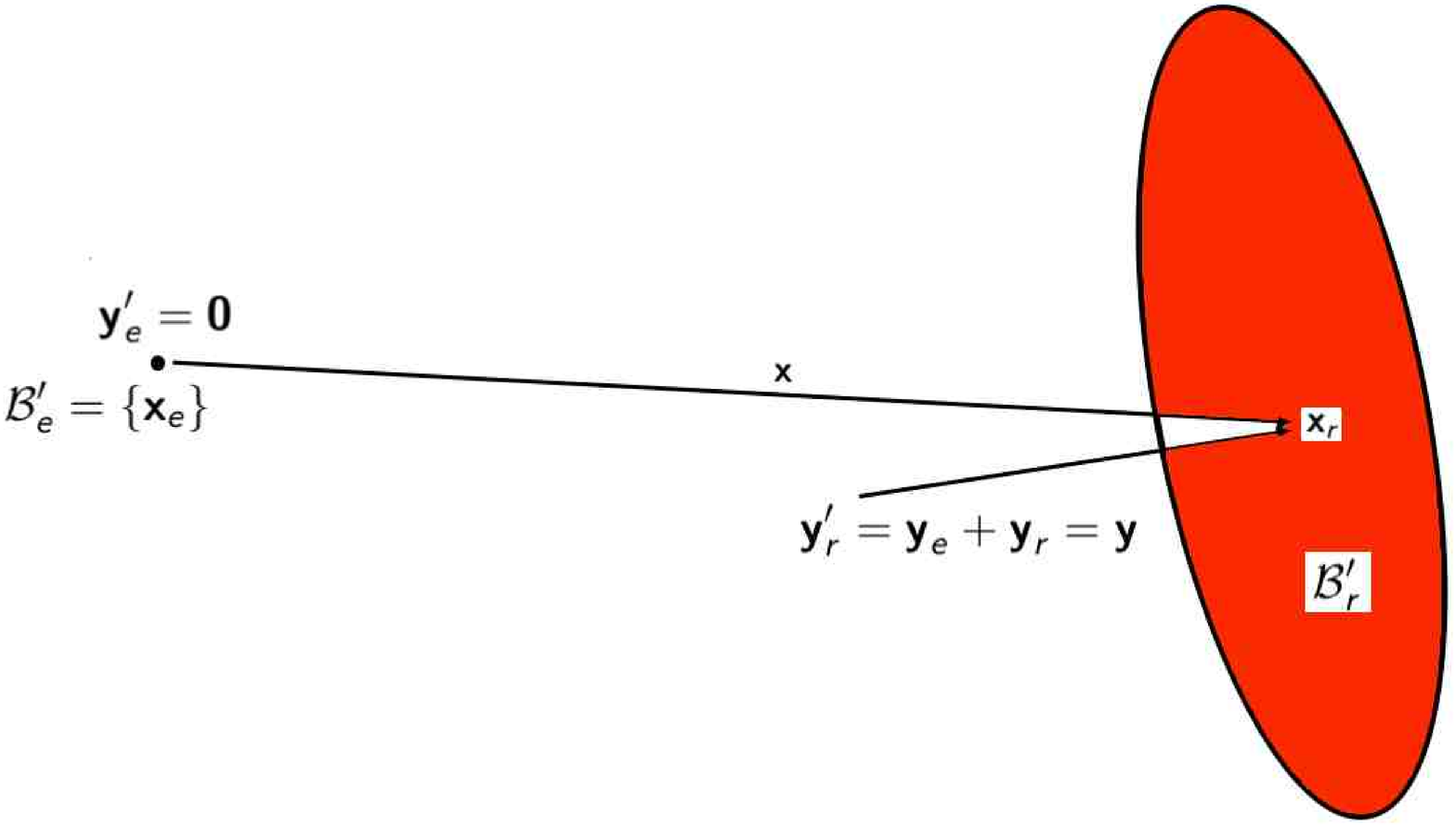}
\caption{Channel $B$ (top) has an idealized event receiver, while channel $C$ (bottom) has an idealized event emitter. Both are equivalent to channel $A$ of Figure 1. Since point emitters and receivers are \sl omnidirectional, \rm channel $B$ needs a larger emitting antenna than channel $A$ to make up for the lost directivity of the receiver, and channel $C$ needs a larger receiving antenna than channel $A$  to make up for the lost directivity of the emitter.}
\label{fig2}
\end{center}
\end{figure}

\section{Bandwidths}

Even though our analysis is in the time domain, it is possible to assign  \sl bandwidths \rm to the emitting antenna, the receiving antenna, and the communication channel as a whole. We define these to be the \sl reciprocals of the shortest pulse that can be emitted, received and exchanged, respectively. \rm The bandwidth of the emitting and receiving antennas are defined as
\begin{align*}
B_e=\frac1{T_e\00}=\frac1{s_e-a_e},\qq B_r=\frac1{T_r\00}=\frac1{s_r-a_r},
\end{align*}
and the \sl effective channel bandwidth \rm is, by \eq{TT},
\begin{align*}
B=\frac1{T\00}=\frac1{s-a}\le \frac1{T_r\00+T_e\00}<\min\{B_e,\ B_r\},
\end{align*}
where $a=|\3y|=|\3y_r+\3y_e|\le a_r+a_e$ is the \sl effective channel aperture. \rm

\section*{Acknowledgements}
I thank Dr. Arje Nachman of AFOSR for his sustained support of my work, most recently through Grant  \#FA9550-04-1-0139.

\end {document}

%% file: GKmacros.tex
\def\={\equiv} 
 
\def\pl{\partial}

\newcommand{\bib}{\bibitem}

\newcommand{\nt}{\notag}
\newcommand{\ci}{\cite}
\newcommand{\lab}{\label}

\newcommand{\eq}{\eqref}

\newcommand{\hr}[1]{\href{#1}{\tt{#1}}}

\newcommand{\lp}{\left(}
\newcommand{\rp}{ \right)}

\newcommand{\LB}{\left\lbrace}
\newcommand{\RB}{\right\rbrace}

\newcommand{\harr}[1]{\smash{\mathop{\hbox to .5in{\ \rightarrowfill\ }}
      \limits^{#1}}}

\newcommand{\0}[1]{{(#1)}}

\newcommand{\2}[1]{{\tilde #1}}
\newcommand{\3}[1]{{\boldsymbol #1}}

\newcommand{\bh}[1]{{\boldsymbol{\hat #1}}}

\newcommand{\6}[1]{_{\scriptscriptstyle#1}}
\newcommand{\7}[1]{{\bar#1}}

\newcommand{\9}[1]{^{\,\scriptscriptstyle#1}}

\def\d{\delta} 
\def\e{\varepsilon}

\def\h{\eta}

\def\o{\omega} 
\def\p{\pi} 

\def\q{\theta} 
 
\def\r{\rho}

\def\t{\tau} 
 
\def\x{\xi}
 
\def\z{\zeta}

\def\D{\Delta}

\def\Q{\Theta}

\newcommand{\hb}[1]{{\ \text{#1}\ }}

\newcommand{\db}{{\,{\rm d}\kern-.9ex {^-}}\!}
\newcommand{\dir}{{\pl\kern-1.2ex {/}}}

\newcommand{\app}{\approx} 
\newcommand{\cc}[1]{{{\mathbb C\hskip.5pt}^{#1}}}

\newcommand{\ie}{{\it i.e., }}

\newcommand{\im}{{\,\rm Im}\ }  
\newcommand{\imp}{\ \Rightarrow\ }
\newcommand{\inv}{^{-1}}

\newcommand{\ir}{\int_{-\infty}^\infty}  

\newcommand{\lra}{\leftrightarrow}

\newcommand{\plra}{\pl^{\kern-1.25ex^\lra}}
\newcommand{\qq}{\quad} 
\newcommand{\qqq}{\qquad} 
\newcommand{\re}{{\,\rm Re}\  }   
\newcommand{\rr}[1]{{{\mathbb R}^{#1}}}

\newcommand{\sgn}{{\,\rm Sgn \,}}

\newcommand{\sr}{\sqrt}

\newcommand{\supp}{{\rm supp \,}}
\newcommand{\sv}[1]{\vskip#1ex}

\def\XXint#1#2#3{{\setbox0=\hbox{$#1{#2#3}{\int}$}
     \vcenter{\hbox{$#2#3$}}\kern-.5\wd0}}

\def\HB{\hfill\break}